# Mutual Photoluminescence Quenching and Photovoltaic Effect in Large-Area Single-Layer MoS$_2$-Polymer Heterojunctions

**Authors:** Tejas A. Shastry,[1,3,+] Itamar Balla,[1,+] Hadallia Bergeron,[1] Samuel H. Amsterdam,[2,3] Tobin J. Marks,[1,2,3] and Mark C. Hersam[1,2,3,*]

[1]Department of Materials Science and Engineering, Northwestern University, Evanston, IL 60208, United States.

[2]Department of Chemistry, Northwestern University, Evanston, IL 60208, United States.

[3]Argonne-Northwestern Solar Energy Research Center, Northwestern University, Evanston, IL 60208, United States.

[+]These authors contributed equally.

*Correspondence to: m-hersam@northwestern.edu

## ABSTRACT

Two-dimensional transition metal dichalcogenides (TMDCs) have recently attracted attention due to their superlative optical and electronic properties. In particular, their extraordinary optical absorption and semiconducting band gap have enabled demonstrations of photovoltaic response from heterostructures composed of TMDCs and other organic or inorganic materials. However, these early studies were limited to devices at the micrometer scale and/or failed to exploit the unique optical absorption properties of single-layer TMDCs. Here we present an experimental realization of a large-area type-II photovoltaic heterojunction using single-layer molybdenum disulfide (MoS$_2$) as the primary absorber, by coupling it to the organic π-donor polymer PTB7. This TMDC-polymer heterojunction exhibits photoluminescence intensity that is tunable as a function of the thickness of the polymer layer, ultimately enabling complete quenching of the TMDC photoluminescence. The strong optical absorption in the TMDC-polymer heterojunction produces an internal quantum efficiency exceeding 40% for an overall cell thickness of less than 20 nm, resulting in exceptional current density per absorbing thickness in comparison to other organic and inorganic solar cells. Furthermore, this work provides new insight into the recombination processes in type-II TMDC-polymer heterojunctions and thus provides quantitative guidance to ongoing efforts to realize efficient TMDC-based solar cells.

**TOC Figure**

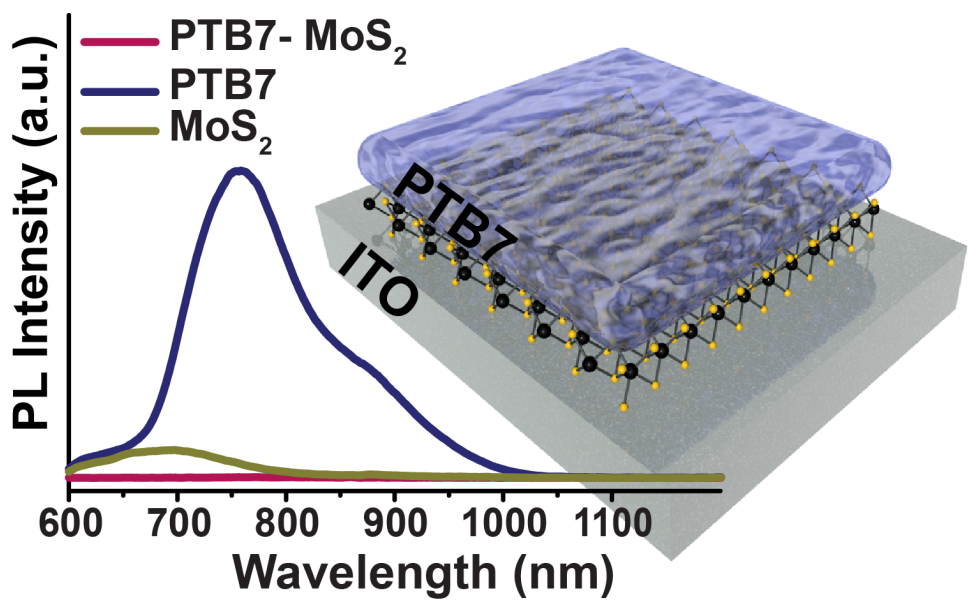

The favorable optoelectronic properties of TMDCs[1-4] have recently enabled demonstrations of transistors and photovoltaics with promising device metrics.[5-19] In particular, solar cells utilizing single-layer TMDCs show exceptional performance per thickness that exceeds many organic and inorganic thin-film photovoltaics.[7, 9] However, these efforts have thus far focused on TMDC-TMDC junctions, where complete photoluminescence quenching is not achieved, thereby limiting the overall current density.[14] Additionally, these early TMDC photovoltaics have been demonstrated only for micrometer scale devices, leaving open questions concerning the viability of integrating ultrathin TMDCs with other thin-film photovoltaic technologies. While large-area TMDC layers have been interfaced to conventional inorganic semiconductors, such as silicon or gallium arsenide,[20, 21] these devices rely on the thick inorganic layer for light absorption and therefore fail to exploit the exceptional performance per thickness that TMDCs offer.

In this work we utilize a mixed-dimensional system that features a π-donor polymer with zero-dimensional (*i.e.*, molecular orbital) electronic structure, poly[[4,8-bis[(2-ethylhexyl)oxy]benzo[1,2-b:4,5-b']dithiophene-2,6-diyl][3-fluoro-2-[(2-ethylhexyl)carbonyl]thieno[3,4-b]thiophenediyl]] (PTB7), to form a type-II photovoltaic heterojunction with single-layer, two-dimensional $MoS_2$ as the acceptor. This scheme allows proper tuning of the PTB7 donor layer thickness to achieve complete quenching of the $MoS_2$ photoluminescence, yielding exceptionally high internal quantum efficiency and current density per thickness that exceeds all other known TMDC and thin-film photovoltaics. The acceptor-induced current in these solar cells is also many times higher than typical organic donor-acceptor systems, establishing $MoS_2$ as the most efficient acceptor to date on a per thickness basis. These insights will inform future development of ultrathin mixed-dimensional solar cells that appropriately couple two-dimensional TMDCs with inorganic and organic materials of higher or lower dimensionality.

**RESULTS AND DISCUSSION**

The structure of single-layer $MoS_2$ is shown in **Figure 1a**.[1] $MoS_2$ was grown over large areas on glass substrates using a previously described chemical vapor deposition (CVD) process.[22, 23] Atomic force microscopy (AFM, **Figure 1b**) of the as-grown CVD $MoS_2$ films reveals thicknesses consistent with monolayer growth,[23] and optical microscopy (**Figures 1c, S1,** and **Table S1**) suggests that only ~20% of the as-grown material consists of bilayer or multilayer regions. A 20 cm$^{-1}$ separation between the $E_{2g}$ and $A_{1g}$ Raman modes in **Figure 1d** and strong photoluminescence (**Figure 1e**) further confirm the presence of large-area single-layer $MoS_2$.[24] An AFM analysis of the $MoS_2$ layer in **Figure S2** shows large, micrometer scale grains constituting the continuous monolayer film, in contrast to the smaller nanometer-scale grain size (*i.e.*, high grain boundary density) exhibited in previous growths of large-area $MoS_2$ monolayers.[25, 26] It has been shown previously that carrier recombination in ultrathin TMDCs is dominated by surface defects, such as grain boundaries.[27] In particular, several studies have shown that grain boundaries in CVD $MoS_2$ can degrade electronic mobility and significantly quench photoluminescence.[26, 28-30] For these reasons, the lower grain boundary density of our continous monolayer $MoS_2$ enables efficient charge transport as is required for photovoltaic applications.

**Mutual Photoluminescence Quenching from a Mixed-Dimensional System**

**Figure 2a** depicts the chemical structure of PTB7, which absorbs in the visible portion of the electromagnetic spectrum (**Figure 2b**). The anticipated type-II alignment between the PTB7 highest occupied molecular orbital (HOMO) and lowest unoccupied molecular orbital (LUMO) levels and the conduction and valence band edges of $MoS_2$ is shown in **Figure 2c**.[13, 31] To verify this energy level alignment, the photoluminescence of a 16 nm thick PTB7 film, a transferred CVD-grown $MoS_2$ film, and a PTB7-$MoS_2$ bilayer heterojunction are provided in **Figure 2d**. The PTB7-$MoS_2$ bilayer heterojunction was formed *via* the scheme in **Figure 3**, in which the CVD-grown $MoS_2$ film on $SiO_2$ is transferred to a glass or indium tin oxide (ITO) substrate as previously described[29,32,33] and then spin-overcoated with PTB7. Strong mutual photoluminescence quenching of both PTB7 and $MoS_2$ is observed in the heterojunction, with minimal PTB7 photoluminescence remaining (**Figure 2d**, inset). Mutual photoluminescence quenching is indicative of charge transfer and the formation of a type-II PTB7-$MoS_2$ heterojunction.[9] In **Figure 2e** (and the linear-scale plot in **Figure S3**), the evolution of $MoS_2$ photoluminescence quenching as a function of PTB7 thicknesses is presented. The tunable thickness of PTB7 enables us to design a photovoltaic junction that extracts all photogenerated charge from the TMDC, in contrast to previous TMDC-TMDC studies where altering the thickness was more difficult and fundamentally altered the electronic properties of the materials.[14] A residual amount of $MoS_2$ photoluminescence is present for a 10 nm thickness of PTB7, while 16 nm thick PTB7 causes complete quenching, with a residual amount of PTB7 photoluminescence. For a 20 nm thickness of PTB7, the PTB7 exhibits strong photoluminescence. Therefore, to ensure complete quenching of the $MoS_2$ photoluminescence while maintaining the thinnest overall active layer, we chose 16 nm thick PTB7 for the fabrication of solar cells.

**$MoS_2$-PTB7 Heterojunction Solar Cells**

Photovoltaic cells were fabricated by evaporating a $MoO_x$ hole transport layer and a silver cathode onto a $MoS_2$-PTB7 bilayer formed on ITO (**Figure 3h, 4a**) with the energy level diagram shown in **Figure 4b**. The photovoltaic performance of the resulting PTB7-$MoS_2$ heterojunction cell under 1 sun (100 mW/cm$^2$, AM 1.5) illumination with a device area of 1.6 mm$^2$ is presented in **Figure 4c** along with the corresponding PTB7-only and $MoS_2$-only control devices. In all cases, the cell active layer is less than 20 nm in thickness. Note that the PTB7-only and $MoS_2$-only devices generate no photovoltage, indicative of resistive behavior from a single material, with no observable Schottky barrier-induced voltage between the contacts and $MoS_2$.[17, 34] In marked contrast, the PTB7-$MoS_2$ bilayer heterojunction exhibits an open-circuit voltage ($V_{oc}$) of 0.21 V, short-circuit current density ($J_{sc}$) of 1.98 mA/cm$^2$, a fill factor (FF) of 24%, and a power conversion efficiency (PCE) of 0.1%. The $J_{sc}$ was calculated by integrating the external quantum efficiency (EQE) spectrum shown in **Figure 4d** against the solar spectrum after correction for lamp spectral mismatch. The EQE shows strong absorption from the single-layer $MoS_2$ optical gap and excitonic transitions as well as a relatively small contribution from the PTB7 layer (97% and 3%, respectively, as evaluated in **Figure S4**), evidenced by the EQE beyond the $MoS_2$ absorption edge. Note that light absorption at wavelengths shorter than 325 nm is not efficiently converted to current due to the absorption of ITO and glass. To date, this device constitutes the largest area demonstration of a solar cell utilizing a TMDC as the absorber.[9, 11, 14-16, 18-21, 34, 35] Several demonstrations of single-layer TMDC photovoltaic absorbers reported similar PCEs of 0.1% - 0.5%.[9, 16, 18] However, these studies were conducted only on single-flake devices with areas of ~50 μm$^2$. In contrast, the 1 mm$^2$ area TMDC solar cells reported here exhibit a PCE comparable to previous reports but with a 20,000× larger area.

The large area of the PTB7-MoS$_2$ heterojunction enables the first direct measurement of a single-layer TMDC photovoltaic EQE over a broad wavelength range using a 1 mm$^2$ monochromated light source and a lock-in measurement technique, as is the standard practice for large-area photovoltaics. This approach contrasts previously reported EQE measurements using a lower power micrometer-sized laser spot to illuminate a junction comprising one or more TMDCs.[12, 18] By quantifying the broadband EQE, the internal quantum efficiency (IQE) can be calculated based on the active layer optical absorption.[36] In particular, a previously established method for organic photovoltaics was employed that accounts for reflectance and parasitic absorbances using the optical constants of the constituent layers (**Figure S5**) and a transfer matrix method (**Figure S6**) to determine the active absorption of the PTB7-MoS$_2$ heterojunction of **Figure 4e**.[36] We note that utilizing the absorption in **Figure 4d** rather than the active absorption as the basis for calculating IQE would result in an erroneously inflated value, since it does not take into account the reflectance and parasitic absorbances that reduce the overall IQE value. Dividing the EQE by the active absorption gives the IQE curve shown in **Figure 4e**, with a maximum greater than 40% at the MoS$_2$ optical gap. This value indicates strong absorption by the MoS$_2$ and also suggests room for improvement by reducing recombination to increase charge collection, particularly in the PTB7. The overwhelmingly large contribution of MoS$_2$ absorption to the power extracted from the solar cell indicates that the thickness-normalized power conversion efficiency is exceptionally high. Specifically, a 0.1% PCE for this 1 nm MoS$_2$ film implies that it is performing as well per thickness as champion organic photovoltaics that possess a 10% PCE at 100 nm thickness[31] and twice as well as champion perovskite solar cells that deliver 20% PCE at 500 nm thickness.[37] PTB7-MoS$_2$ solar cells are thus ideally suited for ultrathin, semi-transparent solar cells.

The most noteworthy performance parameter of these solar cells is the current density per thickness. **Figure 5a** compares the overall current density per thickness for the absorbing layers against other photovoltaic technologies, while **Figure 5b** compares the acceptor-induced current density per thickness against other donor-acceptor systems. For **Figure 5a**, the measured current density is divided by the active-layer absorption thickness, calculated as the percentage contribution to current density of each absorbing component multiplied by the layer thickness of that component. For **Figure 5b**, the measured current density is multiplied by the percentage contribution to current density of the acceptor and then divided by the acceptor thickness. In the case of bulk heterojunction active layers, the volume fraction of the acceptor is multiplied by the overall layer thickness to obtain the acceptor thickness. In **Figure 5a**, it is apparent that all TMDC photovoltaics achieve much higher current density per thickness than competing photovoltaic technologies, establishing their promise as ultrathin solar cells.[14, 16, 18, 38-40] Note also that because our system properly achieves complete quenching of the $MoS_2$, our current density per thickness is greater than previous single-flake photovoltaic demonstrations, and moreover at 20,000× larger area.

Analysis of the acceptor-induced current in **Figure 5b** shows even more exceptional results, where $MoS_2$ outperforms other organic acceptors in various organic donor-acceptor systems by orders of magnitude.[38, 41, 42] As thickness is a proxy for volume, materials with the highest efficiency per thickness enable a direct pathway for cost-effective and resource-effective thin-film solar cells. While increasing the thickness of $MoS_2$ to further exploit these results would affect the semiconducting properties of the active layer, we expect that the high thickness-normalized efficiency observed here can inform efforts for high power conversion efficiency TMDC solar cells through light trapping schemes,[43] tandem solar cell architectures, or the use of TMDCs with thickness-invariant properties.[44]

**Recombination in PTB7-MoS$_2$ Heterojunctions**

To further probe the PTB7-MoS$_2$ heterojunction recombination characteristics, J$_{sc}$ and V$_{oc}$ were measured as a function of light intensity in **Figure 6**. In **Figure 6a**, the V$_{oc}$ data are fit to the relation of **Equation 1**,

$$V_{oc} \propto S \ln I \qquad (1)$$

where I is the light intensity and S is the slope of the relation. A slope of 2kT/q or greater (such as observed here) is commonly attributed to trap-assisted recombination dominating over bimolecular recombination.[45] In **Figure 6b**, the light-intensity dependent current J$_{sc}$ is fit to a power-law relation (**Equation 2**), where I is the light intensity, and the exponent α is related to the charge transfer dynamics within the system,

$$J_{sc} \propto I^\alpha \qquad (2)$$

The observed value of α close to 1 suggests that carrier transport is not limited by space charge effects, bimolecular recombination, or variations in mobility.[45] The small thickness of this solar cell coupled with the high carrier mobility of MoS$_2$ likely contributes to the nearly ideal observed carrier transport.[1] In **Figure 6c**, the fill factor is observed to remain fairly constant with increasing light intensity. In addition, the steep dependence of voltage on light intensity leads to an enhancement of power conversion efficiency at higher intensity illumination up to 0.3%. From these data, we conclude that the recombination in PTB7-MoS$_2$ heterojunctions is distinct from other organic solar cells, which show decreasing fill factor with increasing light intensity and higher bimolecular recombination.[45] Furthermore, the linear current dependence, IQE results, and recombination-limited voltage suggest that while carrier transport is desirable in this system, reducing recombination at the MoS$_2$ surface should increase the cell charge extraction and open circuit voltage.

**CONCLUSIONS**

Large-area type-II TMDC-polymer heterojunctions have been fabricated and characterized. Mutual photoluminescence quenching confirms efficient charge transfer between PTB7 and single-layer MoS$_2$ following optical excitation, and the tunable PTB7 layer thickness ensures complete quenching. Bilayer PTB7-MoS$_2$ heterojunction solar cells exhibit high absorption from the optical and excitonic bands of the MoS$_2$ as well as from PTB7. At less than 20 nm total thickness, the PTB7-MoS$_2$ heterojunction solar cells provide a peak EQE exceeding 20% and a peak IQE exceeding 40%, resulting in record current density per thickness compared to all other known photovoltaic technologies. In contrast to classical organic photovoltaic systems, trap-assisted recombination dominates PTB7-MoS$_2$ heterojunctions, suggesting that efficiency improvements can be achieved by reducing surface traps on MoS$_2$. This fundamental understanding is likely to inform future optimization efforts aimed at realizing large-area, ultrathin, and high-efficiency solar cells based on TMDC-polymer heterojunctions.

**METHODS**

**Growth of MoS$_2$.** Monolayer MoS$_2$ was grown on 4 cm × 1 cm 300 nm thick SiO$_2$/Si wafers (SQI Inc.) *via* chemical vapor deposition (CVD). Prior to growth, the substrate was sonicated in acetone and isopropyl alcohol (IPA) for 10 min each and subsequently subjected to an O$_2$ plasma (Harrick Plasma) at ~200 mTorr for 1 min with 10.2 W power applied to the RF coil.

During growth, the substrate was placed in the middle of the hot zone of a 1" diameter quartz tube furnace (Lindberg/Blue). Alumina boats with 15 mg molybdenum trioxide ($MoO_3$, 99.98% trace metal, Sigma-Aldrich) and 150 mg sulfur powder (Sigma-Aldrich) were used as solid sources for the reaction. The $MoO_3$ boat was placed upstream and in close proximity to the substrate, whereas the sulfur was positioned 30 cm upstream of the $MoO_3$ boat with a proportional integral derivative temperature-controlled heating belt. The reaction tube was pumped to ~60 mTorr base pressure and purged with ultrahigh purity Ar gas at 200 sccm. After 10 min, the pressure was increased to 400 Torr, and the tube evacuated to its base pressure. Once purging was completed, the pressure was set to 150 Torr with a constant flow of Ar as carrier gas at 25 sccm for the remainder of the reaction and cooling of the furnace. The furnace was heated from room temperature (RT) to 150 °C in 5 min and soaked at that temperature for 20 min to eliminate residual physisorbed contaminants in the tube and on the substrate. Subsequently, the furnace was brought to 800 °C in 55 min and soaked for additional 20 min at that temperature. Concurrently, the heating belt was set to heat to 50 °C in 5 min and soak for 49 min until the furnace temperature was ~500 °C. After the initial soak, the belt was brought to 150 °C in 22.5 min and soaked for an additional 23 min. Finally, the furnace and heating belt were cooled to RT naturally.

**Transfer of $MoS_2$.** As-grown monolayer $MoS_2$/300nm $SiO_2$/Si samples were spin-coated with 2-3 drops of PMMA ($M_W$~950,000 at a concentration of 4% wt. in anisole, MicroChem) at 3,000 RPM for 60 sec and annealed in air for 10 min at 180 °C. To etch away the $SiO_2$ and delaminate the PMMA/$MoS_2$ thin film from the $SiO_2$/Si substrate, the sample was immersed in 3 M potassium hydroxide (KOH) solution for 12-48 h. Subsequently, a clean glass slide was used to transfer and float the delaminated PMMA/$MoS_2$ thin film in a clean DI water bath. This process was repeated twice to remove excess etchant residue. Next, the film was transferred to the target substrate and annealed for 5 min at 80 °C and 15 min at 150 °C. Once the sample was cooled to room temperature, it was immersed in chloroform (HPLC Plus Sigma) for 24 h to dissolve the PMMA. Finally, the sample was rinsed with IPA and blown dry with $N_2$ gas for further processing.

**Raman Measurements.** A Horiba Scientific XploRA PLUS Raman microscope was used to acquire Raman spectra. Prior to measurement, the instrument was calibrated with the Si band at 520.7 $cm^{-1}$ using a Si(100) wafer. Spectral acquisition was performed under ambient conditions with an excitation laser line of 532 nm (spot size ~1 $\mu m^2$) and incident power of ~1 mW for an acquisition time of ~30 s to avoid heating effects. The Raman signal was collected using a 100× Olympus objective (NA = 0.9) and dispersed by a 1800 grooves/mm grating to a Syncerity CCD detector with a spectral resolution finer than 2 $cm^{-1}$.

**AFM Measurements.** Atomic force microscopy (AFM) experiments were carried out using an Asylum Cypher AFM in tapping and lateral force microscopy modes. Si cantilevers NCHR-W (NanoWorld) with a resonant frequency of ~300 kHz were used for tapping mode imaging. For lateral force microscopy measurements, FMR cantilevers (NanoWorld) with resonant frequency of ~75 kHz and contact force of ~5 nN were used. Scanning in this mode was performed with the fast scan direction perpendicular to the cantilever. The images were taken with a pixel resolution of 1024 × 1024 at a scanning rate of 1~1.5 Hz.

**Polymer Solution Preparation**. Poly[[4,8-bis[(2-ethylhexyl)oxy]benzo[1,2-b:4,5-b']dithiophene-2,6-diyl][3-fluoro-2-[(2-ethylhexyl)carbonyl]thieno[3,4-b]thiophenediyl]] (PTB7) was purchased from Ossila and used as received. PTB7 was dissolved in chlorobenzene at a concentration of 5 mg/mL and stirred for two days at 60 °C to completely dissolve the polymer.

**Absorbance and Photoluminescence Measurements.** Optical absorbance measurements were taken using an Agilent Cary 5000 on films of individual active layer components spin-coated on ITO. Photoluminescence measurements were taken on single-layer or bilayer thin films on glass substrates. For polymer and $MoS_2$ photoluminescence, a 5 nm slit width and 0.5 s integration time were used. An excitation wavelength of 500 nm and 350 nm were used for PTB7 and $MoS_2$ photoluminescence, respectively. Measurements were taken on a Horiba Spectrometer with a low pass filter blocking the excitation light from the silicon emission detector.

**Device Fabrication.** Pre-patterned indium tin oxide (ITO) glass substrates (1 × 2 $in^2$, 15 Ohm/Sq Thin Film Devices) were cleaned by ultrasonic treatment in aqueous detergent (Alconox), deionized water, acetone, and isopropyl alcohol, sequentially. To fabricate photovoltaic devices, $MoS_2$ was transferred onto ITO as described above, and 20 μL of the polymer solution described above was spread across the $MoS_2$ portion of the ITO substrate and spin-coated in an inert $N_2$ atmosphere at 1,000 RPM for 60 s. Next, the films were placed in a glovebox-enclosed thermal evaporator and pumped down to 5 x $10^{-6}$ Torr. Device fabrication was completed by depositing 7.5 nm of $MoO_3$ (Alfa-Aesar, Puratronic 99.9995%) and 100 nm of silver (Lesker). A device area of 1.6 $mm^2$ was defined as the overlap between the patterned ITO, active layer, and silver cathode.

**Device Testing.** For 1 sun measurements, the completed devices were tested under 100 mW/$cm^2$ of calibrated solar simulated light from a Xenon arc lamp source (Newport) with an AM 1.5G filter. For higher illuminations, the source lamp was focused using lenses to a smaller spot size, and the intensity was calibrated using a silicon photovoltaic detector with a known efficiency by comparing the output short circuit current of the device. In all cases, the concentrated illumination size was larger than 1 $cm^2$. External quantum efficiency measurements were performed with a 75 W monochromated light source utilizing a Xenon arc lamp for the visible portion of the spectrum and a tungsten lamp for the near-infrared portion of the spectrum.

Measurements were calibrated with a silicon photodetector. All measurements were taken using devices without encapsulation under ambient conditions.

**Internal Quantum Efficiency (IQE) Calculations.** IQE was calculated by dividing the external quantum efficiency by $1 - R$, where R is the reflectance. The reflectance was calculated using the transfer matrix method using optical constants derived from databases and literature values as presented in the caption of Figure S5.

**Competing financial interests.** The authors declare no competing financial interests.

**Supporting Information Available:** Additional information including supplemental optical images, AFM, and additional measurements. This material is available free of charge *via* the Internet at http://pubs.acs.org/.

**Acknowledgments.** The device fabrication and testing were supported as part of the Argonne-Northwestern Solar Energy Research (ANSER) Center, an Energy Frontier Research Center funded by the U.S. Department of Energy, Office of Science, Basic Energy Sciences under Award DE-SC0001059. The $MoS_2$ CVD growth was supported by the National Institute of Standards and Technology (NIST CHiMaD 70NANB14H012). T.A.S. and S.H.A. acknowledge fellowships from the Patrick G. and Shirley W. Ryan Foundation. H.B. acknowledges support from the NSERC Postgraduate Scholarship-Doctoral Program. The authors kindly thank Xiaolong Liu for valuable discussions.

**Contributions.** T.A.S. and I.B. contributed to this work equally.

**FIGURES**

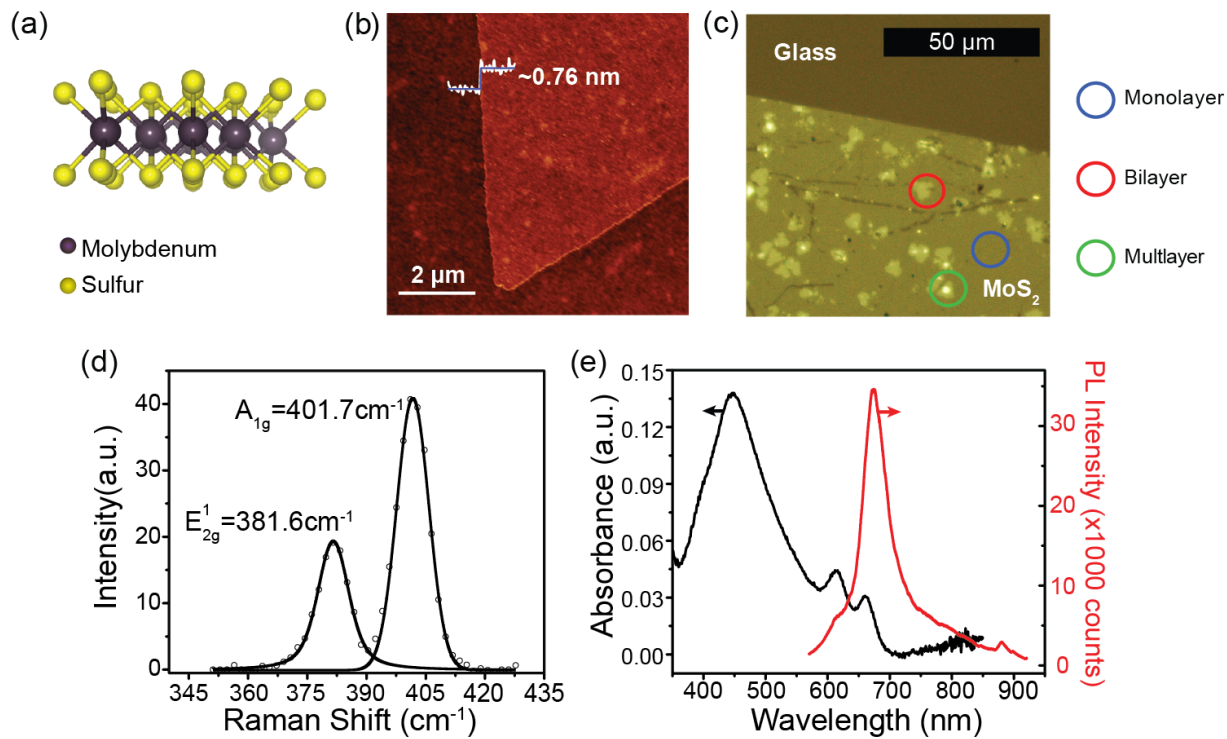

**Figure 1.** Structure and properties of CVD-grown $MoS_2$. (a) Atomic structure of $MoS_2$. (b) Atomic force micrograph of CVD-grown monolayer $MoS_2$ exhibiting a thickness less than 1 nm. (c) Optical micrograph of CVD-grown $MoS_2$ transferred onto glass. The monolayer $MoS_2$ film is visible along with small regions of bilayer and multilayer growth. (d) Raman spectrum of CVD-grown $MoS_2$, exhibiting a peak separation indicative of monolayer $MoS_2$. (e) Optical absorbance and photoluminescence spectra of CVD-grown $MoS_2$ transferred onto glass.

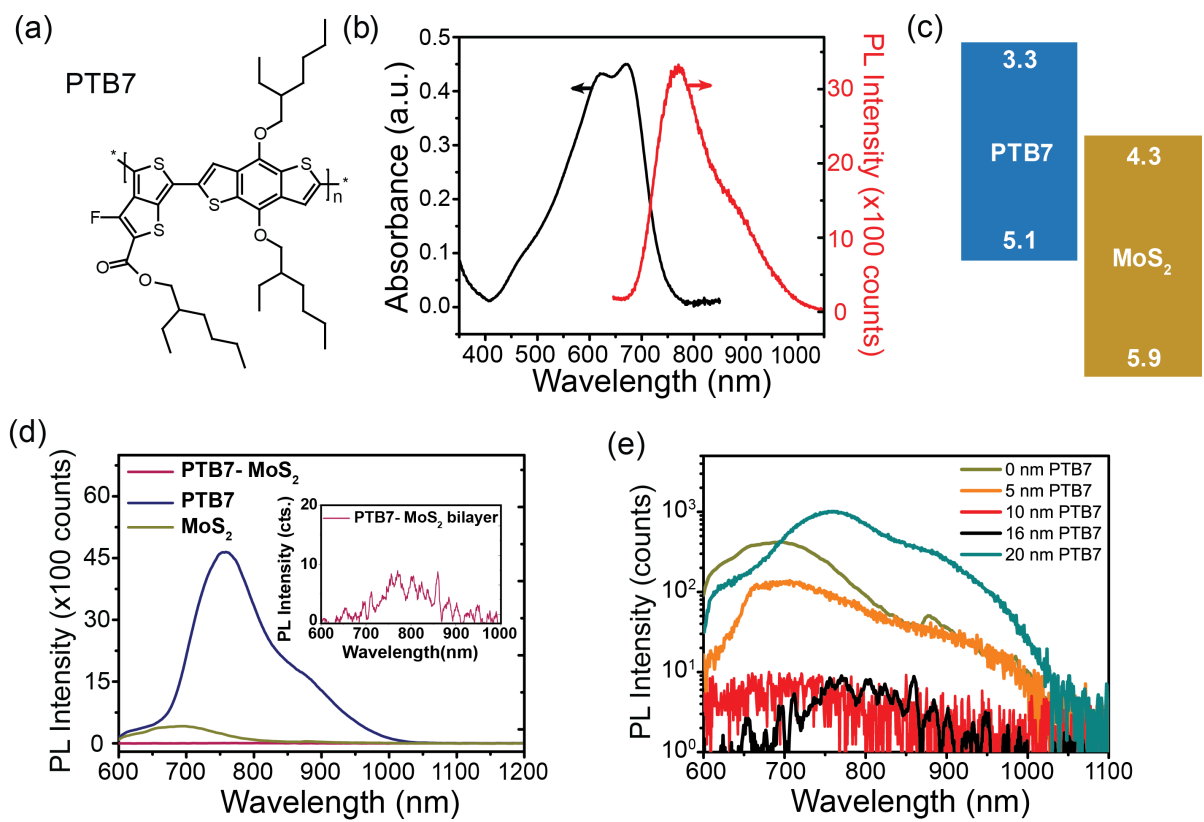

**Figure 2.** Charge transfer between $MoS_2$ and PTB7. (a) Chemical structure of the PTB7 polymer. (b) Optical absorbance and photoluminescence spectra of a spin-coated PTB7 film. (c) Band alignment for PTB7 and $MoS_2$. (d) Photoluminescence spectra for films of PTB7 alone, $MoS_2$ alone, and a PTB7-$MoS_2$ bilayer. The inset shows nearly complete quenching of $MoS_2$ photoluminescence and a small contribution of remaining PTB7 photoluminescence. (e) $MoS_2$ photoluminescence as a function of PTB7 layer thickness.

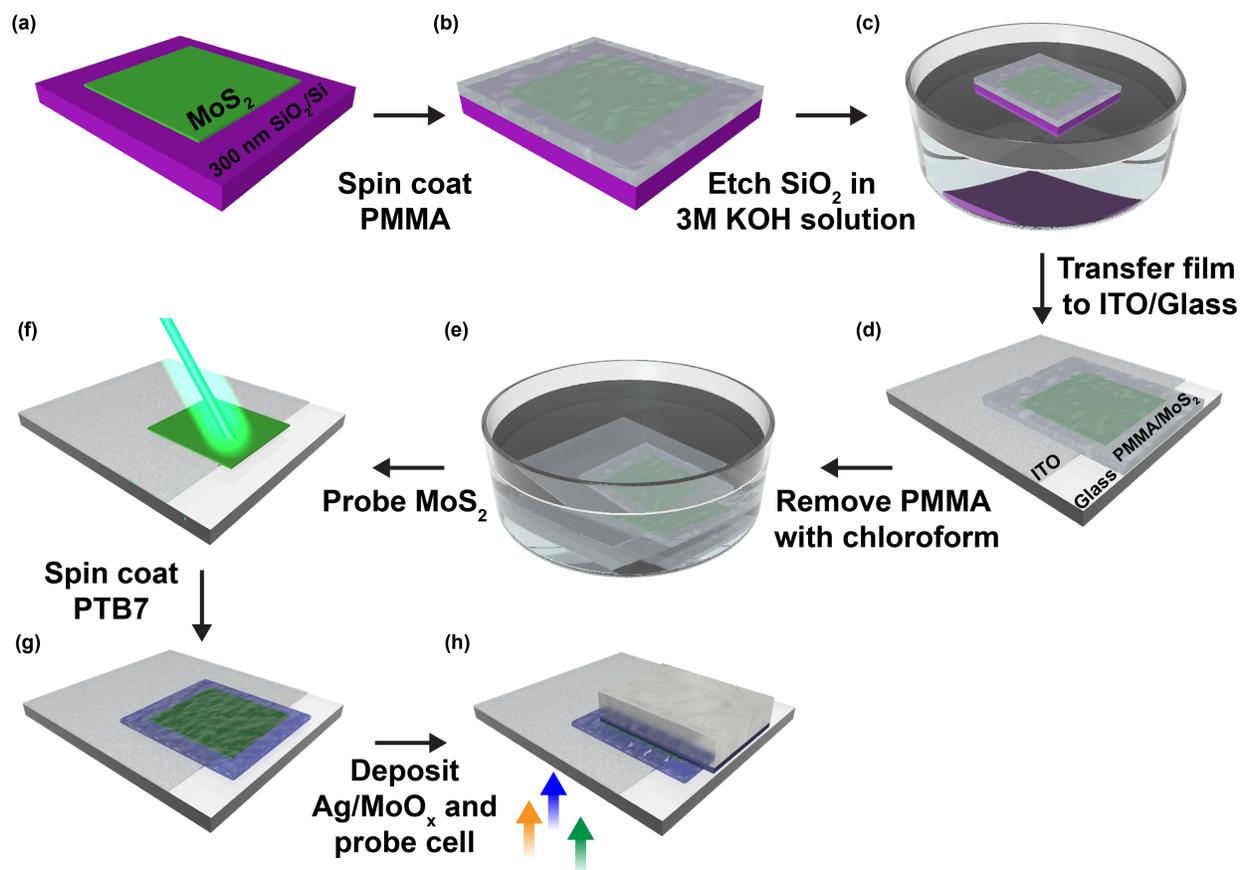

**Figure 3.** Photovoltaic cell fabrication. (a) Monolayer MoS$_2$ grown on 300 nm SiO$_2$/Si. (b)-(e) Transfer of MoS$_2$ to ITO/glass substrate. (f) Probing MoS$_2$ alone before spin coating it with PTB7 (g). (h) Evaporating a MoO$_x$ hole transport layer and a silver cathode onto the MoS$_2$-PTB7 bilayer completes the cell.

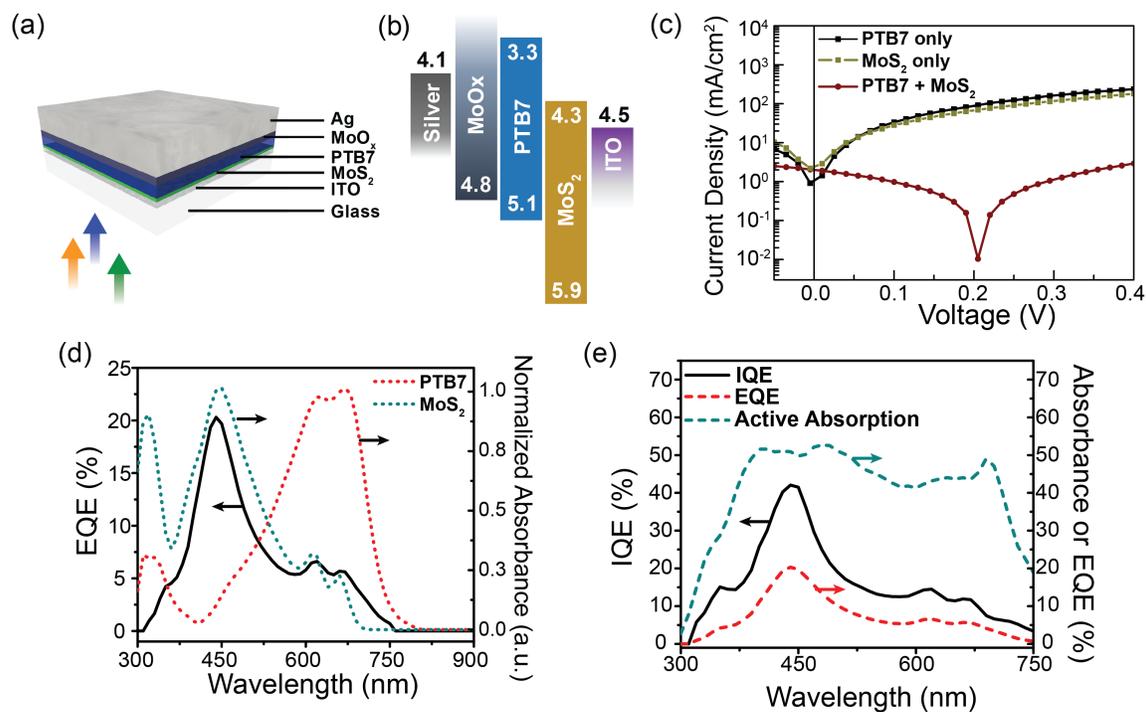

**Figure 4.** MoS$_2$/PTB7 solar cells. (a) Device cross section structure, with a 16 nm thick PTB7 layer. (b) Energy levels of the cell depicted in (a). (c) Current-voltage response of a device with PTB7 only, MoS$_2$ only, and a PTB7/MoS$_2$ bilayer. (d) External quantum efficiency of the solar cell in (c), with the optical absorbance spectra of PTB7 and MoS$_2$ for reference. (e) Calculated internal quantum efficiency of the solar cell.

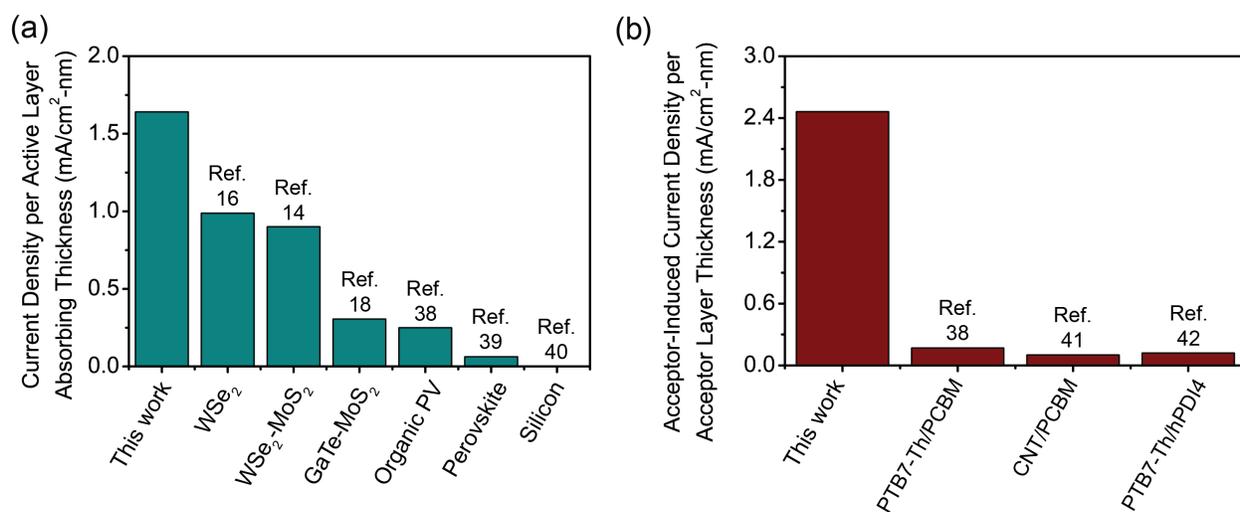

**Figure 5.** Performance per thickness comparison. (a) Overall current density per active layer thickness for this work compared to other TMDC demonstrations and other photovoltaic technologies. (b) Current density attributed to the acceptor per acceptor layer thickness for this work compared to other organic donor-acceptor systems.

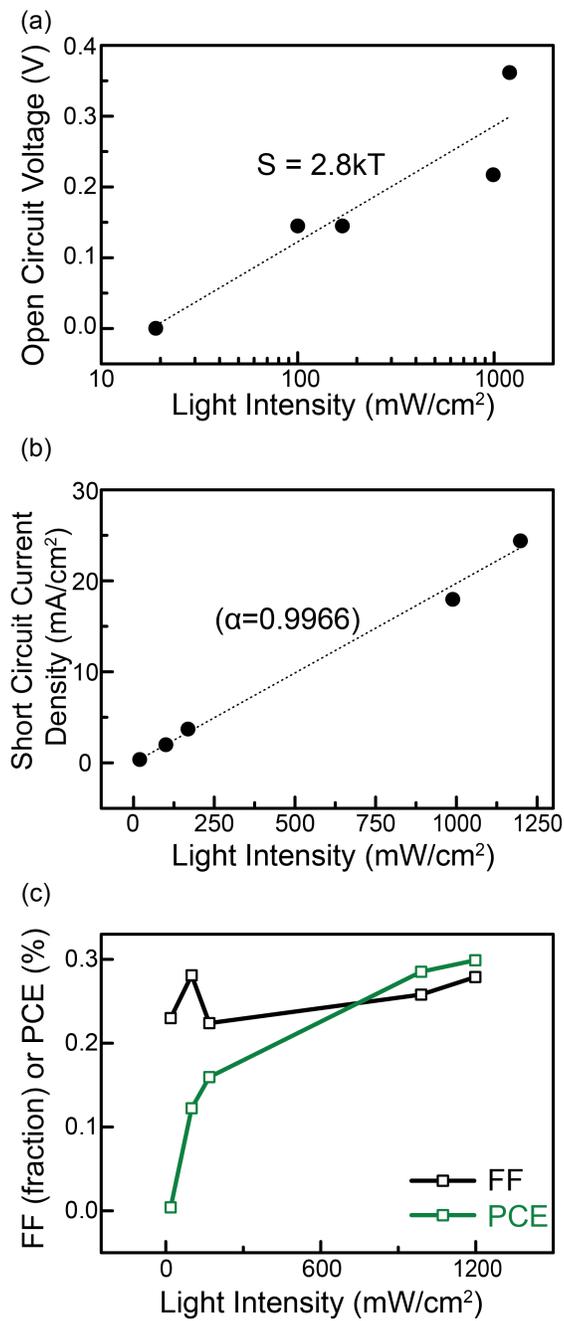

**Figure 6.** Recombination in MoS$_2$/PTB7 solar cells. Light intensity dependence of (a) open circuit voltage, (b) short circuit current density, and (c) fill factor and power conversion efficiency for an MoS$_2$/PTB7 solar cell.